\documentclass[cameraready]{Interspeech}

\title{CAAD: Contrastive Audio-Aware Distillation for Efficient Speech Language Models}
\usepackage{hyperref}

\author[affiliation={1}]{Chun-Wei}{Chen}
\author[affiliation={2}]{Tzu-Quan}{Lin}
\author[affiliation={2}]{Ke-Han}{Lu}
\author[affiliation={2}]{Wei-Ping}{Huang}
\author[correspondingauthor, affiliation={1,2,3},]{Hung-Yi}{Lee}

\address{
    $^1$Graduate Institute of Electrical Engineering, National Taiwan University \\
    $^2$Graduate Institute of Communication Engineering, National Taiwan University,\\
    $^3$NTU Artificial Intelligence Center of Research Excellence (NTU AI-CoRE), Taiwan
}

\email{r14921061@ntu.edu.tw hungyilee@ntu.edu.tw}


\usepackage{comment}
\usepackage{multirow}
\usepackage{cite}

\begin{document}

\maketitle

\begin{abstract}
Speech Language Models achieve reasoning capabilities, but are often hindered by massive parameter counts and a tendency to prioritize linguistic priors over acoustic features. While contrastive decoding enhances grounding by contrasting audio-aware and text-only logits, it increases inference latency. We propose Contrastive Audio-Aware Distillation (CAAD), a framework that internalizes the teacher's contrastive reasoning into the student model's weights. To overcome the high computational training overhead in the dual-path token-by-token contrastive distillation process, we introduce a synchronized teacher-forcing strategy. Anchored by unified Pseudo-Ground Truths, this mechanism enables simultaneous full-sequence generation of the teacher's contrastive distributions, allowing student to distill the audio-aware signal efficiently. Overall, CAAD yields a $\sim$8\% relative gain over standard knowledge distillation on Dynamic-SUPERB and successfully reduces linguistic bias in MCR-BENCH.
\end{abstract}

\section{Introduction}
The convergence of large-scale linguistic and acoustic modeling has catalyzed the development of Speech Language Models (SLMs)\cite{lu2024desta,lu2025desta2,lu2025developing,xu2025qwen25omnitechnicalreport,tang2023salmonn,ding2025kimi,zhang2023speechgpt,arora2025landscape, yang2024building,cui2025recent}, capable of instruction following and cross-modal reasoning. However, the massive parameter count of these foundation models poses significant challenges for low latency \cite{cui2025recent}. Consequently, some research has shifted to Knowledge Distillation for model compression.

A fundamental limitation of applying standard Knowledge Distillation (Std. KD) to multimodal SLMs is that it inadvertently transfers the teacher's own modal biases. Large teacher SLMs sometimes struggle to maintain sufficient focus on input audio, as their strong internal linguistic priors can easily ignore acoustic evidence \cite{wu2025language}. Because the current method uses standard Knowledge Distillation frameworks \cite{nouriborji2025efficient} to train the student model minimizing the divergence from the teacher's final output distribution, the student copies these linguistic priors. Consequently, the distilled model learns to lean on the teacher model's inertia rather than actively grounding its predictions in the rich information of the audio modality.

Contrastive Decoding (CD) \cite{hsu2025reducing} offers a powerful remedy for linguistic biases in Speech Language Models (SLMs). At its core, CD isolates the desired signal by comparing the simultaneous positive and negative generation paths (dual-path) at each decoding step. By penalizing reliance on the negative path, CD enforces stronger grounding. While we find that this dual-path approach consistently improves models, it inherently doubles inference latency because it requires the computation of two forward passes per token.

A natural solution is to distill this contrastive reasoning into a single-path student model. However, transferring token-by-token decoding logits to a student is computationally expensive~\cite{agarwal2024policy,zhao2026self}. Because standard CD relies on autoregressive dual-path generation, calculating the positive and negative path targets token-by-token also breaks the parallelization typically utilized during training, resulting in massive computational overhead.

To overcome these bottlenecks, we propose \textbf{Contrastive Audio-Aware Distillation (CAAD)}\footnote{\href{https://github.com/ChenWils/Contrastive_Audio-Aware_Distillation.git}{https://github.com/ChenWils/Contrastive\_Audio-Aware\_Distillation.git}}. Our core innovation is an efficient contrastive training objective specifically designed for Speech Language Models (SLMs) that internalizes the teacher's acoustic grounding abilities without the expensive training. To make this viable, we utilize a synchronized teacher-forcing strategy, anchoring both positive and negative teacher paths to a unified Pseudo-Ground Truth (Pseudo-GT) that uses text metadata derived from the audio in DeSTA framework. By doing so, we enable simultaneous full-sequence logit generation. This unified Pseudo-GT allows the student to efficiently learn the grounding benefits of CD into a standard single-path model while accelerating training.

In summary, our contributions are summarized as follows:
\begin{itemize}
    \item \textbf{Contrastive SLM Distillation:} We introduce \textbf{CAAD}, a novel objective that internalizes contrastive reasoning into student weights. This isolates the audio-aware signal and eliminates the high latency of dual-path inference.
    \item \textbf{Synchronized Teacher-Forcing Strategy:} We propose a synchronized teacher-forcing strategy anchored by \textbf{Pseudo-Ground Truths}, enabling aligned generation of teacher contrastive distributions to maintain full training parallelization.
    \item \textbf{Empirical Performance:} Evaluated on \textbf{Dynamic-SUPERB} \cite{huang2024dynamic} and \textbf{MCR-BENCH}\cite{wang2025audio}, CAAD effectively mitigates linguistic bias. The student model consistently outperforms \textbf{standard knowledge distillation} and \textbf{contrastive decoding} in test time, while surpassing the greedy decode performance of teacher model in paralinguistic tasks.
\end{itemize}

\section{Related Work}
\subsection{Modality Bias in Speech Language Models}
A critical bottleneck in Speech Language Models (SLMs) is modality bias, where the model's powerful linguistic backbone creates an uneven power struggle between text and audio. Rather than treating both as equal partners, the model consistently trusts its internal linguistic priors first, suppressing speech features during inference \cite{wu2025language}. This structural hierarchy ensures that acoustic evidence is neglected, relegating speech signals to a secondary status \cite{kuan2024understanding, leng2024curse}. This bias is deeply ingrained across diverse architectures, causing models to default to familiar modality even when sensory inputs provide clear contradictory evidence \cite{zheng2025mllms}.

This imbalance manifests during the process of deciding which source of information to trust. Under conditions of cross-modal conflict, SLMs frequently ignore the acoustic signal in favor of linguistic instructions \cite{billa2026audio}. Recent benchmarks such as MCR-BENCH confirm that when audio and text disagree, models almost invariably follow the text, signaling a failure in true multimodal reasoning \cite{wang2025audio}. This linguistic prior is particularly detrimental in intent detection, where models rely on transcripts even when the primary intent is only identifiable via paralinguistic cues, leading to performance degradation in audio-aware queries \cite{mullick2025text}.

To counteract this modality bias, recent research has leveraged contrastive learning to enforce stronger cross-modal alignment during inference. While contrastive learning has been used extensively for distillation \cite{chang2024colld, zhu2025ckd}, they can apply to SLMs to penalize linguistic prior\cite{lin2026contrastivedecodingenhanceslarge}. For instance, Audio-Aware Decoding \cite{hsu2025reducing} addresses this during inference by contrasting multimodal output distributions against those from missing modalities to neutralize the linguistic prior. However, while these interventions successfully improve cross-modal grounding, they double generation latency. This motivates the need for more efficient alignment strategies that resolve modality bias without sacrificing real-time performance.

\section{Methodology}
\begin{figure}[t] 
    \centering 
    \includegraphics[width=0.47\textwidth]{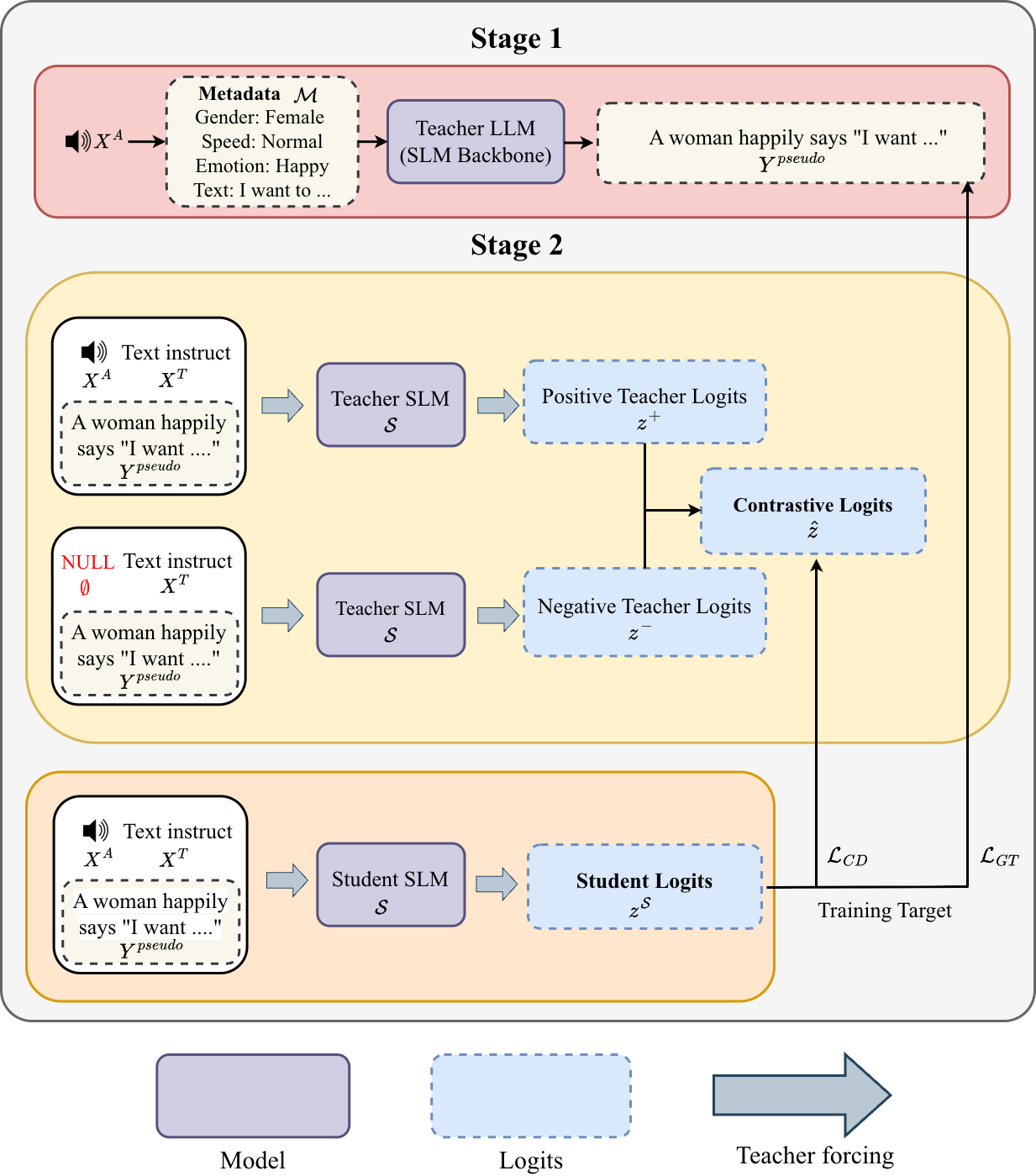}
    \vspace{-5mm}
    \caption{Overview of the two-stage CAAD framework where Stage 1 generates a Pseudo-GT anchor and Stage 2 applies dual-path contrastive distillation to train the student model.}
    \vspace{-5mm}
    \label{fig:Pipeline} 
\end{figure}

A fundamental challenge in distilling multimodal contrastive decoding is the expensive computational overhead of generating dual-path targets autoregressively. Since standard teacher forcing relies on a single shared sequence to accelerate training via full-sequence parallelization, simultaneously computing a teacher's positive (audio-aware) and negative (text-only) distributions requires a fixed anchor. To address this, we propose the \textbf{Contrastive Audio-Aware Distillation (CAAD)} training framework in \figurename~\ref{fig:Pipeline} to solves this bottleneck in two stages.

\subsection{Stage 1: Pseudo-GT Generation via Metadata Anchor}

In Stage 1, we generate a \textit{Pseudo-Ground Truth (Pseudo-GT)} to serve as the unified anchor for both teacher paths.
Following the DeSTA \cite{lu2024desta,lu2025desta2,lu2025developing} framework, we construct a \textit{Pseudo-GT} dataset to synchronize the distillation process. In our primary formulation, given an audio input $X^A$, we first extract text metadata $M$ that include gender, emotion, acoustic environment, and so on. Large Language Model backbone of the SLMs generates a structurally dense and descriptive text sequence \textit{Pseudo-GT} $Y^{pseudo}$ strictly conditioned on $M$. This resulting sequence $Y^{pseudo} = \{y_1, y_2, \dots, y_L\}$ serves as the fixed anchor for both teacher passes during Stage 2 distillation. We also explore an alternative baseline for generating $Y^{pseudo}$ directly from the continuous audio modality. Our empirical analysis in Section 5.2 ablation studies demonstrates that text-based metadata provides a higher-fidelity anchor for the synchronized teacher-forcing strategy.

\subsection{Stage 2: Contrastive Audio-Aware Distillation}
In Stage 2, we utilize $Y^{pseudo}$ in a synchronized teacher-forcing strategy, optimizing a student model to efficiently capture the contrastive, audio-aware shift logit of the frozen teacher $\mathcal{T}$.

\begin{itemize}
    \item \textbf{Positive Path (Audio-aware):} Captures the joint distribution of audio ($X^A$) and text instructions ($X^T$).
    \begin{equation}
        z_{t}^+ = \mathcal{T}(X^A, X^T, y_{<t}^{pseudo})
    \end{equation}
    
    \item \textbf{Negative Path (Text-only):} Captures the teacher’s linguistic priors by masking the audio input ($\emptyset$).
    \begin{equation}
        z_{t}^- = \mathcal{T}(\emptyset, X^T, y_{<t}^{pseudo})
    \end{equation}
\end{itemize}

We define the \textit{Audio-Aware Target} $\hat{z}$ by extrapolating the logit space away from the negative path. We introduce a guidance scaling factor $\alpha \ge 0$ to amplify the audio-aware contribution:
\begin{equation}
    \hat{z}_t = (1+\alpha) \cdot z_{t}^+ - \alpha \cdot z_{t}^-
\end{equation}
By subtracting the negative path components, we penalize tokens heavily supported by teacher's linguistic priors and reward those uniquely activated by the audio-aware signal $X^A$.

\subsection{Optimization Objective}
The student model $\mathcal{S}$ minimizes a hybrid objective that aligns its parameters with the teacher's audio-aware signal, maintaining linguistic fluency while overcoming linguistic priors.

\begin{itemize}
    \item \textbf{Contrastive Distillation Loss ($\mathcal{L}_{CD}$):} We minimize the Kullback-Leibler (KL) divergence between the student's output distribution and the sharpened teacher target to internalize the audio-aware shifts:
    \begin{equation}
        \mathcal{L}_{CD} = \frac{1}{L}\sum_{t=1}^{L}\tau^2 \cdot \text{KL}\left( \sigma(\hat{z}_t / \tau) \parallel \sigma(z^\mathcal{S}_t / \tau) \right)
    \end{equation}
    where $z^\mathcal{S}_t$ are the student logits, $\sigma$ is the softmax function, and $\tau$ is the temperature hyperparameter.
    
    \item \textbf{Pseudo-GT Supervision Loss ($\mathcal{L}_{GT}$):} To maintain linguistic fluency and prevent the model from deviating into ungrammatical regions of the logit space during distillation, we apply a standard Cross-Entropy (CE) loss against the Pseudo-GT tokens:
    \begin{equation}
        \mathcal{L}_{GT} = \frac{1}{L}\sum_{t=1}^{L}\text{CrossEntropy}(z^\mathcal{S}_t, y_t^{pseudo})
    \end{equation}

     \item \textbf{Total Training Loss:} The final objective is a weighted combination of the distillation signal and the supervised ground truth:
    \begin{equation}
        \mathcal{L}_{total} = \lambda \mathcal{L}_{CD} + (1 - \lambda) \mathcal{L}_{GT}
    \end{equation}
    where $\lambda \in [0,1]$ is a balancing hyperparameter. 
\end{itemize}

\begin{table*}[t]
\caption{Performance comparison across all Dynamic-SUPERB categories and MCR-BENCH conflict resolution tasks. We evaluate the DeSTA2 across various distillation and decoding method. \textbf{Bold} denotes the best student performance; \underline{Underline} indicates the student outperforming the greedy decoding teacher.}
\vspace{-2mm}
\label{tab:full_desta2_results}
\centering
\resizebox{\textwidth}{!}{
\begin{tabular}{@{}llcccccc|ccccc@{}}
\toprule
 & & \multicolumn{6}{c}{\textbf{Dynamic-SUPERB(\%)} $\uparrow$} & \multicolumn{5}{c}{\textbf{MCR-BENCH} (\%)} \\
\cmidrule(lr){3-8} \cmidrule(l){9-13}
\textbf{Model Size} & \textbf{Decoding / Distill Mode} & \textbf{CON} & \textbf{SEM} & \textbf{PAR} & \textbf{DEG} & \textbf{SPK} & \textbf{ALL} & \textbf{$Acc_{neu}$} & \textbf{$Acc_{fth}$} & \textbf{$Acc_{adv}$} & \textbf{$Acc_{irr}$} & \textbf{Shift} $\downarrow$ \\ \midrule
Teacher (8B) & Greedy Decode & 79.41 & 59.42 & 43.14 & 51.63 & 42.50 & 56.78 & 3.90 & 98.60 & 1.10 & 41.20 & 97.37 \\
Teacher (8B) & CD  & 81.72 & 62.92 & 52.14 & 59.73 & 44.57 & 61.79 & 11.20 & 51.40 & 15.00 & 41.80 & 83.96 \\ \midrule
Student (3B) & Greedy Decode & 54.45 & 49.42 & 32.78 & 39.84 & 22.92 & 41.02 & 1.40 & \textbf{97.40} & 1.00 & 34.00 & 90.65 \\
Student (3B) & CD  & 40.13 & 43.57 & 29.42 & 38.52 & 20.21 & 35.80 & 1.00 & 56.60 & 7.40 & 26.90 & 87.50 \\
Student (3B) & Std. KD  & 65.72 & 58.42 & 43.35 & 46.42 & \textbf{36.14} & 50.40 & 41.00 & 96.90 & 0.50 & 40.9 & 100 \\
\textbf{Student (3B)} & \textbf{CAAD (Ours)} & \textbf{73.86}& \underline{\textbf{60.57}} & \underline{\textbf{51.35}} & \textbf{49.23} & 35.00 & \textbf{54.44} & \underline{\textbf{45.90}} & 82.80 & \underline{\textbf{11.80}} & \underline{\textbf{45.50}} & \underline{\textbf{79.03}} \\ \bottomrule
\end{tabular}
}
\vspace{-2mm}
\end{table*}

\section{Experiment}
\subsection{Training Dataset}

To ensure distillation and robust acoustic generalization, we utilize expressive speech instruction-following dataset established by the DeSTA2, which consolidates diverse benchmarks including AccentDB\cite{ahamad2020accentdb}, DailyTalk \cite{lee2023dailytalk}, IEMOCAP \cite{busso2008iemocap}, PromptTTS \cite{guo2023prompttts}, VCTK \cite{yamagishi2019cstr}, and VoxCeleb \cite{nagrani17_interspeech}.

Beyond the original labels, the corpus employs specialized pre-trained models to extract paralinguistic features, including gender, emotion state \cite{ma2024emotion2vec}, pitch, speaking rate, and environmental acoustics such as signal-to-noise ratio (SNR) and C50 \cite{lavechin2023brouhaha}. In total, samples are annotated with 12 distinct attributes covering speaker identity, prosody, and acoustic environment.

For the unified anchor, we rely on pseudo-GT generated by the teacher's LLM backbone. Following DeSTA protocol, Llama3-8B-Instruct \cite{grattafiori2024llama} serves as stage-1 teacher for the DeSTA 3B student, with temperature and top-$p$ set to 1.

\subsection{Model Architectures and Training Configurations}

To demonstrate the architecture of our proposed CAAD framework, we train our approach in DeSTA2. 
The teacher model is instantiated using the standard DeSTA2 architecture with a Llama-3.2-8B foundation. To construct the student model, we substitute the Large Language Model backbone with the smaller Llama-3.2-3B. Following the standard DeSTA2 training protocol, we keep the core LLM parameters strictly frozen during distillation. We optimize only the Q-Former modality adapter to project acoustic features into the LLM embedding space. This strategy limits the trainable parameters to 32M, significantly reducing the computational overhead. The model was optimized using FusedAdam with a learning rate of $1 \times 10^{-4}$ and a cosine schedule. $\mathcal{L}_{total}$ is weighted by $\lambda = 0.7$ and $\tau = 2.0$ for $\mathcal{L}_{CD}$. Training is executed on an RTX A6000 GPU, totaling 70 hours.

\subsection{Evaluation Setup}
To rigorously assess the student model's ability to balance instruction-following with acoustic perception, we employ the \textbf{Dynamic-SUPERB} benchmark \cite{huang2024dynamic}. This framework is specifically designed to evaluate instruction-tuned speech language models on unseen tasks, making it the standard for measuring generalization in SLMs. We systematically categorize the Dynamic-SUPERB tasks into five distinct dimensions: \textbf{Content} (CON), \textbf{Semantic} (SEM), \textbf{Paralinguistic} (PAR), \textbf{Degradation} (DEG), and \textbf{Speaker} (SPK). In general, tasks within the CON and SEM dimensions can be predominantly resolved utilizing linguistic and linguistic priors. Conversely, the PAR, DEG, and SPK dimensions demand acoustic perception, requiring the model to leverage non-verbal paralinguistic cues and environmental information that extend beyond the transcribed text.

To evaluate the model's ability to mitigate linguistic priors, we employ \textbf{MCR-BENCH}~\cite{wang2025audio}, a framework specifically designed for modality conflict resolution. We focus our evaluation on the Speech Emotion Recognition subset derived from MELD \cite{poria2019meld}. Unlike standard datasets, MCR-BENCH introduces scenarios where the linguistic prompt and the acoustic signal may provide misleading or irrelevant information. We evaluate performance across four key accuracy metrics: \textbf{Neutral} ($Acc_{neu}$), which presents the original question without additional context; \textbf{Faithful} ($Acc_{fth}$), featuring accurate audio descriptions; \textbf{Adversarial} ($Acc_{adv}$), containing deliberately misleading text that contradicts the audio; and \textbf{Irrelevant} ($Acc_{irr}$), using irrelevant or nonsensical descriptions. To quantify the impact of linguistic prior shifts, we measure the percentage of correct answers that flip to incorrect when exposed to misleading text. We calculate the \textbf{Shift} numerical value as follows:
\begin{equation}
    Shift = \frac{\Delta_{c \to i}}{N_{neu}}
\end{equation}
where \textbf{$\Delta_{c \to i}$} is the number of instances correctly predicted in the Neutral setting but failed in the Adversarial setting, and \textbf{$N_{neu}$} is the total number of correct predictions in the Neutral setting. A lower Shift value indicates the model is more robust to misleading text and relies more on the acoustic signal. By analyzing these metrics and value, we can quantify a model's reliance on acoustic evidence versus its linguistic priors.

\section{Results}
To evaluate the impact of our distillation framework, 
we compare our CAAD method against different baselines:
(1) \textbf{Greedy Decoding}, representing the model's vanilla baseline where the model selects the token with the highest probability at each step; 
(2) Contrastive Decoding (CD), a test-time method that amplifies acoustic signals by subtracting text-only logits from audio-aware logits. We propose an optimal hyperparameter configuration for this approach; and
(3) \textbf{Standard KD (Std. KD)}, which utilizes traditional KL-divergence for knowledge transfer in logits without using the contrastive audio-aware signal.
By comparing these, we can isolate whether performance gains come from simple knowledge transfer or our CAAD method.

\subsection{Main Performance Comparison}

The results in Table \ref{tab:full_desta2_results} demonstrate that CAAD  serves as a more robust training objective compared to Std. KD and CD methods. While Std. KD often causes the student to inherit linguistic priors, CAAD successfully isolates and amplifies the audio-aware signal, leading to a consistent performance uplift across Dynamic-SUPERB categories. 
Notably, CAAD demonstrates superior stability compared to CD. Whereas applying CD at test time triggers a performance collapse in student model, CAAD successfully internalizes the contrastive logic directly into the model weights. This approach ensures structural stability, empowering the 3B student to outperform the teacher's vanilla greedy baseline in semantics and paralinguistics. Although it falls short of the CD teacher model, the CAAD student prioritizes audio-aware signals better than its original source model.

On the MCR-BENCH conflict resolution tasks, CAAD demonstrates the ability to mitigate linguistic bias in favor of the audio-aware signal. Greedy decoding in both teacher and student model tend to over-rely on linguistic instructions, leading to failure when audio contradicts or is irrelevant to the prompt. CAAD effectively breaks this reliance, achieving a much more sophisticated balance between instruction following and audio-aware signal utilization. By successfully focusing on the audio rather than simply predicting based on text patterns, CAAD significantly outperforming the teacher and Std. KD baselines in neutral ($Acc_{neu}$), adversarial ($Acc_{adv}$), and irrelevant ($Acc_{irr}$) tasks. Besides, the lower Shift score establishes CAAD as a more reliable architecture for multi-modal reasoning independent of linguistic priors

\begin{table}[t]
\caption{Ablation of the contrastive distillation weight $\alpha$ in DeSTA2. $\alpha=0.0$ represents Std. KD. We report the Average (ALL) score on Dynamic-SUPERB and the Shift value on MCR-BENCH to demonstrate the impact of the contrastive signal on performance and linguistic priors.}
\vspace{-2mm}
\label{tab:ablation_alpha_short}
\centering
\resizebox{\columnwidth}{!}{
\begin{tabular}{@{}lcc@{}}
\toprule
\textbf{Config ($\alpha$)} & \textbf{Dynamic-SUPERB (ALL)} $\uparrow$ & \textbf{MCR-BENCH (Shift)} $\downarrow$ \\ \midrule
$\alpha = 0.0$ (Std. KD)   & 50.40   & 100.00                                  \\
$\alpha = 0.5$ (CAAD)      & 53.63   & 98.51                                   \\
$\alpha = 1.0$ (CAAD)      & \textbf{55.00}   & 93.78                          \\
$\alpha = 2.0$ (CAAD)      & 54.44    & \textbf{79.03}                         \\ \bottomrule
\end{tabular}
}
\vspace{-5mm}
\end{table}

\subsection{Ablation Study}

\textbf{Sensitivity to Contrastive Weight ($\alpha$):} Table \ref{tab:ablation_alpha_short} illustrates the critical role of the contrastive scaling factor $\alpha$ in balancing general performance with modality bias. We find that all tested contrastive configurations ($\alpha > 0$) consistently outperform the Std. KD ($\alpha = 0.0$), confirming that the contrastive signal is essential for enhancing performance. While $\alpha=1.0$ yields the highest average (ALL) score of 55.00 on Dynamic-SUPERB, any increment in the contrastive weight progressively mitigates the inherent linguistic bias of the model. This is most evident in the MCR-BENCH Shift value, which drops from a maximum of 100.00 at the baseline to 79.03 at $\alpha=2.0$. This trend demonstrates that while moderate weighting optimizes acoustic accuracy, higher values $\alpha$ successfully forcing the model to focus on audio-aware signals over linguistic priors and achieving a more robust multi-modal integration.

\textbf{Effect of the Pseudo-Ground Truth Anchor:} To validate the necessity of using high-quality text metadata $M$ to generate the fixed anchor $Y^{pseudo}$, we conduct an ablation study comparing it against a baseline generated directly from continuous audio $X^A$. Table \ref{tab:ablation_sync_transposed} demonstrates that CAAD consistently improves general performance and mitigates modality bias regardless of the synchronization modality used. However, deriving the Pseudo-GT from text metadata provides a higher-fidelity anchor compared to raw audio inputs. While Audio Synchronization yields Dynamic-SUPERB score of 49.83 and MCR-BENCH Shift value to 94.46, Metadata Synchronization under the CAAD framework further optimizes these metrics and values. Specifically, it boosts the performance score to 54.44 and enhances de-biasing, effectively reducing the MCR-BENCH Shift value to 79.03. We attribute this to the model's superior internal representation and generation stability when conditioned on structured metadata rather than continuous audio embeddings. These findings confirm that while CAAD provides the audio-aware signal, the metadata-driven $Y^{pseudo}$ is essential for a synchronized teacher-forcing strategy.

\begin{table}[t]
\caption{Ablation of Pseudo-GT synchronization strategies in DeSTA2. We report the Average (ALL) score on Dynamic-SUPERB and the Shift value on MCR-BENCH.}
\vspace{-2mm}
\label{tab:ablation_sync_transposed}
\centering
\resizebox{\columnwidth}{!}{
\begin{tabular}{@{}lcccc@{}}
\toprule
 & \multicolumn{2}{c}{\textbf{Audio Sync.}} & \multicolumn{2}{c}{\textbf{Metadata Sync. (Ours)}} \\
\cmidrule(lr){2-3} \cmidrule(l){4-5}
\textbf{Metric / Value} & \textbf{Std. KD} & \textbf{CAAD} & \textbf{Std. KD} & \textbf{CAAD} \\ \midrule
Dynamic-SUPERB (ALL) $\uparrow$ & 46.80 & 49.83 & 50.40 & \textbf{54.44} \\
MCR-BENCH (Shift) $\downarrow$  & 99.45 & 94.46 & 100.00 & \textbf{79.03} \\ \bottomrule
\end{tabular}
}
\vspace{-5mm}
\end{table}

\section{Conclusion}
In this work, we presented \textbf{Contrastive Audio-Aware Distillation (CAAD)}, distillation framework designed to compress Speech Language Models while mitigating linguistic priors. Through the synchronized teacher-forcing strategy anchored by metadata-driven pseudo-ground truths, CAAD effectively bridges the performance gap between student and teacher models. Evaluations on the Dynamic-SUPERB benchmark demonstrate that CAAD outperforms standard knowledge distillation and contrastive decoding, even allowing student to surpass teacher in some paralinguistic tasks. More importantly, results from MCR-BENCH conflict resolution tasks confirm that CAAD mitigates linguistic bias in speech language models. By successfully focus on the acoustic signal, CAAD achieve a superior balance between instruction following and acoustic perception, establishing a more robust and reliable foundation for multi-modal reasoning in small scale SLMs.

\section{Limitations}
The efficacy of knowledge distillation often depends on the performance gap between the teacher and student models. When employing a highly optimized student architecture, such as Qwen2.5-Omni 3B as a student model, the potential for further gain through distillation may be marginal. Consequently, as our proposed method relies on distillation frameworks, its effectiveness may be inherently bounded by the pre-existing proficiency of small language models.

\section{Acknowledgement}
This work was supported in part by the Ministry of Education (MOE), Taiwan, through the NTU Artificial Intelligence Center of Research Excellence (NTU AI-CoRE) under the framework of the Taiwan Centers of Excellence in Artificial Intelligence project. The authors would like to express their sincere gratitude to the MOE for its financial assistance, which greatly facilitated this research. Furthermore, we are deeply grateful to the National Center for High-performance Computing (NCHC) of the National Applied Research Laboratories (NARLabs), Taiwan, for generously providing the vital computational infrastructure and storage resources that made the extensive experiments in this study possible.

\section{Generative AI Use Disclosure}
 The authors maintain full responsibility for the research design, experimental execution, data analysis, and the final reported results. Generative AI tools are employed solely for linguistic refinement and polishing of the manuscript. These AI tools do not contribute to the substantive scientific content or intellectual framework of the study.

\bibliographystyle{IEEEtran}
\bibliography{mybib}

\end{document}